\documentclass[a4paper,12pt]{article}

\usepackage[english]{babel}
\usepackage[utf8x]{inputenc}
\usepackage[T1]{fontenc}
\usepackage{authblk}
\usepackage[dvipdfmx]{graphicx}
\usepackage{natbib}
\usepackage{here}
\usepackage{setspace}
\usepackage{multirow}
\usepackage{lmodern}
\usepackage{amsmath, amssymb}
\usepackage{bm}
\usepackage{pdflscape}
\usepackage{authblk}
\usepackage{color}
\usepackage{siunitx}

\usepackage{booktabs}
\usepackage{afterpage}
\usepackage{textcomp}
\usepackage{float}
\usepackage{enumerate}
\usepackage{rotating}

\usepackage[margin=1in]{geometry}

\usepackage{amsmath}
\usepackage{graphicx}
\usepackage[colorlinks=true, allcolors=blue]{hyperref}

\usepackage{caption}
\usepackage{subcaption}

\usepackage{lineno}

\begin{document}

\begin{titlepage}
    \begin{center}
        \vspace*{1cm}
        \large
        \textbf{Characterizing functional relationships between anthropogenic and biological sounds: A western New York state soundscape case study} \\
         \normalsize
          \vspace{5mm}
          \textbf{Running Title}: Characterizing biological and anthropogenic sounds \\
         \vspace{5mm}
         Jeffrey W. Doser\textsuperscript{1}, Kristina M. Hannam\textsuperscript{2}, Andrew O. Finley\textsuperscript{1, 3} \\
         \vspace{5mm}
    \end{center} 
     \small
         \textsuperscript{1}Department of Forestry, Michigan State University, East Lansing,
 MI, USA \\
         \textsuperscript{2}Department of Biology, State University of New York (SUNY) Geneseo, Geneseo, New York, USA \\
         \textsuperscript{3}Department of Geography, Environment, and Spatial Sciences, Michigan State University, East Lansing, Michigan, USA \\
 \noindent \textbf{Corresponding Author}: Jeffrey W. Doser, telephone: (585) 683-4170; email: doserjef@msu.edu; ORCID ID: 0000-0002-8950-9895
\doublespacing

\section*{Abstract}

\subsection*{Context}

Roads are a widespread feature of landscapes worldwide, and road traffic sound potentially makes nearby habitat unsuitable for acoustically communicating organisms. It is important to understand the influence of roads at the soundscape level to mitigate negative impacts of road sound on individual species as well as subsequent effects on the surrounding landscape.

\subsection*{Objectives}

We seek to characterize the relationship between anthropogenic and biological sounds in western New York and assess the extent to which available traffic data explains variability in anthropogenic noise.

\subsection*{Methods}

Recordings were obtained in the spring of 2016 at 18 sites throughout western New York. We used the Welch Power Spectral Density (PSD) at low frequencies (0.5-2 kHz) to represent anthropogenic noise and PSD values at higher frequencies (2-11 kHz) to represent biological sound. Relationships were modeled using a novel two-stage hierarchical Bayesian model utilizing beta regression and basis splines.

\subsection*{Results}

Model results and map predictions illustrate that anthropogenic noise and biological sound have an inverse relationship, and anthropogenic noise is greatest in close proximity to high traffic volume roads. The predictions have large uncertainty, resulting from the temporal coarseness of public road data used as a proxy for traffic sound.

\subsection*{Conclusions}

Results suggest that finer temporal resolution traffic sound data, such as crowd-sourced time-indexed traffic data from geographic positioning systems, might better account for observed temporal changes in the soundscape. The use of such data, in combination with the proposed modeling framework, could have important implications for the development of sound management policies.

\vspace{5mm}

\noindent \textbf{Keywords}: Bayesian, bioacoustics, ecoacoustics, road effect, soundscape ecology, anthropogenic noise

\subsection*{Acknowledgements}

We thank the Genesee Valley Conservancy, Monroe Community College, Jeff Over, and the Geneseo Roemer Arboretum for permission to use their land as recording locations. JWD was supported by the Geneseo Foundation Undergraduate Summer Fellowship and National Science Foundation (NSF) EF-1253225. AOF was supported by NSF DMS-1513481, EF-1241874, and EF-1253225. All authors declare that they have no conflict of interest.

\end{titlepage}

\doublespacing

\section*{Introduction}

Roads are a widespread feature of most landscapes worldwide, with road networks growing dramatically in the past 100 years. In the United States alone, there are over 6.3 million kilometers of public roads, most of those (80\%) found in rural areas \citep{forman2003}.  Nowhere in the United States is very far from a road, with the farthest straight line distance from a road in the lower 48 states being a spot in Wyoming 21 miles from the nearest road \citep{projectRemote}.  Since 1970, the traffic on US roads has at least tripled to almost 5 trillion vehicle kilometers traveled per year \citep{barber:2010}. Habitat fragmentation caused by roads is detrimental to wildlife due to direct mortality via wildlife-vehicle collisions, exposure to pollutants, and exposure to sound from road traffic \citep{barber:2011, parris2008, mcclure2013, snow2018}. Thus, while roads alter habitats and landscapes structurally, impacts of roads on animal diversity and abundance can also be impacted by altered acoustic environments \citep{katti2004, mcclure2013}.

Acoustic space, or the soundscape, is an essential resource for both terrestrial and marine animals \citep{pijanowski-2-2011, farina2018}.  Animals utilize the auditory spectrum for a variety of functions, including reproduction \citep{mcgregor2005}, predation and to warn of danger \citep{templeton2006, marler2004, sloan:2008, ridley2007}, and to find food \citep{rice1982, knudsen1979, neuweiler1989}. The sounds organisms produce are collectively called biological sound, which combine together with abiotic sounds from the earth, like wind and rushing water, and human-produced sounds (anthropogenic noise) to form the soundscape \citep{pijanowski-2-2011}. Road sound may be the most pervasive form of anthropogenic noise impacting natural habitats and contributes sound with particular characteristics to the soundscapes of those habitats. Sound from a road is a linear rather than a point source \citep{katti2004}, the sound from traffic tends to be low frequency (typically below 2 kHz) and high amplitude, and the timing of road sound in some places can vary greatly over time (e.g., rush hour peaks) and depend on traffic load \citep{slabbekoorn2008}.  

Traffic sound and other sources of anthropogenic noise have created soundscapes with novel acoustic characteristics in which acoustically communicating animals send and receive signals. High amounts of anthropogenic noise reduce the perception of biologically important sound \citep{barber:2010} and are thought to have negative effects on both cognitive processes \citep{Potvin2017} as well as behavior \citep{brumm2005}. Traffic sound often masks auditory signals, limiting or preventing senders and receivers from communicating effectively, a phenomenon that is  well-documented \citep{brumm2005, patricelli:2006}, particularly for birds and frogs. Traffic sound was shown to cause physiological stress and impair breeding behavior in multiple frog populations throughout the world \citep{tennessen2014}, and similar effects have been demonstrated in birds \citep{ortega2012, warren2006}. Some bird species are able to respond to anthropogenic noise by adapting characteristics of their song to overcome masking. A study on Song Sparrows (\textit{Melospiza melodia}), found a positive correlation between the minimum frequency of a male’s song and the amplitude of low-frequency background noise in a range of urban environments \citep{wood2006}, suggesting the organisms are attempting to adapt to increased low frequency sound by changing the pitch of their songs to overcome masking. However, responses to anthropogenic noise are species specific, and not all species are able to change signal frequency or amplitude in a short term response to increased sound in the environment \citep{patricelli:2006, oberweger2001, brackenbury1978}. Further, even organisms that are able to adapt their signals may suffer from reduced fitness \citep{phillips2018}, suggesting anthropogenic noise can have negative effects even on the species that change their signals in response to increasing sound \citep{patricelli:2006}. An alternative response to anthropogenic noise is for species to avoid habitats where it impacts the soundscapes, a conclusion drawn from tests of the ``phantom road'' effect \citep{mcclure2013} and observations of changes in species abundances near roads \citep{fahrig2009}.

Without proper management of anthropogenic noise, the negative impacts could cause changes in species composition with potentially far reaching effects on the ecosystem. Thus, it is necessary to analyze the relationship between biological sound and anthropogenic noise and understand how it changes across temporal and spatial gradients in order to accurately predict how anthropogenic noise will influence the species comprising the biological sound. More specifically, the identification of anthropogenic noise ``hot spots'' in space and time will allow natural resource managers and others to pinpoint the times and locations in which human sound should be mitigated to maintain the integrity of local ecosystems \citep{ortega2012}. A potential first step in addressing these complex problems is developing a model that quantifies the relationship between biological sound and prominent sources of anthropogenic noise, such as road sound.

Ecoacoustics researchers \citep{gage2017, seur2008} have developed a number of acoustic indices, such as the Acoustic Complexity Index (ACI) \citep{aci}, Acoustic Diversity Index (ADI) \citep{villa2011}, Bioacoustic Index (BI) \citep{boelman2007}, and Normalized Difference Soundscape Index (NDSI) \citep{ndsi} to quantify soundscapes and understand how biological sound relates to anthropogenic noise. NDSI was developed to compare the relative amounts of anthropogenic noise and biological sound within an environment \citep{ndsi} and has been shown to correlate well with landscape characteristics despite its relative simplicity \citep{fuller2015}, and is thus a reasonable measure to further characterize the relationship between anthropogenic noise and biological sound across different spatio-temporal gradients. The NDSI is built using the Welch power spectral density (PSD) \citep{welch1967} for 1 kHz frequency bins within the recording.

The NDSI and PSD are useful tools for tracking spatio-temporal changes in soundscapes \citep{mullet2016, pijanowski-2-2011}. However, soundscape data present some unique challenges that potentially cannot be addressed using these indices and simple statistical models. For example, the data used in this study are multivariate (partitioned into frequencies associated with anthropogenic and natural sounds), compositional (frequency ranges sum to total sound at a given location and time), non-Gaussian, and are highly correlated across time. Soundscape data are also typically sparsely sampled in space, highly correlated across space, and often comprise high-dimensional continuous time series for short time intervals with large intervening time gaps. While contemporary statistical literature offers modeling theory for such data complexities \citep{clarkTextbook, hooten}, applied methodology and software are not yet available in the field of ecoacoustics. Application of statistical models that can address these complexities can lead to improved inference and prediction on important soundscape variables across space and time, quantified by statistically valid estimates of uncertainty. 

In this study, we propose a hierarchical Bayesian modeling approach to assess the spatial distribution of biological sound and anthropogenic noise in western New York soundscapes in relation to roads and traffic density. Hierarchical Bayesian models (HBMs) offer an intuitive framework to decompose complex ecological problems into logical parts (data, process, and parameters) \citep{berliner1996, cressie09}. The framework is ideal for drawing inference about soundscapes as it can accommodate high-dimensional, multivariate, compositional data with time and space dependence. Specifically, our objectives are to: 1) characterize the functional relationship between anthropogenic noise and biological sound; 2) assess the extent to which available traffic data explains variability in anthropogenic noise; 3) develop a methodology to deliver statistically valid maps of anthropogenic noise and biological sound that reflect the relationship identified in Objective 1 with accompanying uncertainty quantification. 

\section*{Materials and Methods}

\subsection*{Study Location and Data Collection}

Recording sites were located in nine forest patches in western New York. The forest plots included two arboretums managed by colleges (State University of New York at Geneseo and Monroe Community College), two forest plots owned and managed by the Genesee Valley Conservancy, two forest plots in the town of Brighton, and one forest plot each in the towns of Avon, Conesus, and Rush. Forest plot locations were chosen along an urban-rural gradient to provide soundscapes with varying exposure to levels of anthropogenic noise. 

We obtained recordings at two locations within each of the nine forest plots, resulting in 18 total recording sites. The two locations within each site were chosen to provide recordings of soundscapes at varying distances to the nearest major road. From May-June 2016 we obtained three 30 minute recordings at each recording site in the morning (between 6-8am), afternoon (between 12-2pm), and evening (between 6-9pm), resulting in a total of 54 30 minute recordings. All recordings were taken on weekdays and specific recording times within the specified sampling periods were chosen randomly. We recorded in stereo at a sampling rate of 44.1 kHz using a Song Meter SM4 from Wildlife Acoustics \citep{wildlifeAcoustics} mounted on a tripod one meter above the ground using the built-in omnidirectional microphones with a sensitivity of $-36 \pm 4$dB.  We discarded the last minute of each 30 minute recording as a result of extraneous sound. Each 29 minute recording was broken up into 29 consecutive one-minute sound bites, resulting in a total of $n = 18 \text{ sites } \times\; 3 \text{ times per day } \times\; 29 \text{ sound bites} = 1566$ observations. We recorded on days with similar weather conditions during which birds are known to communicate (i.e., no rain, minimal wind) to minimize any influence of weather on the observed soundscape patterns. 

We used basic scatter plots and standard regression functions in R \citep{bates2015} to assess the influence of the day of the recording and the time of a morning recording from sunrise on the relationship between biological sound and anthropogenic noise. We did not find large amounts of variability in this relationship across the differing factor levels, and so we do not include them as covariates in the subsequent modeling framework. 

\subsection*{Soundscape Metrics}

Recordings were manually analyzed using a sampling scheme for rapid soundscape analysis \citep{Bobryk2016} to determine the most common acoustically communicating organisms in the soundscapes. The analysis revealed the soundscapes contained varying amounts of anthropogenic noise depending on the site and were dominated by avian species, the most common species including American Robin (\textit{Turdus migratorius}), Cedar Waxwing (\textit{Bombycilla cedrorum}), Yellow Warbler (\textit{Setophaga petechia}), Red-eyed vireo (\textit{Vireo olivaceus}), and Eastern Wood-Pewee (\textit{Contopus virens}).

\begin{figure}
    \centering
    \includegraphics[width = 16cm, height = 9cm]{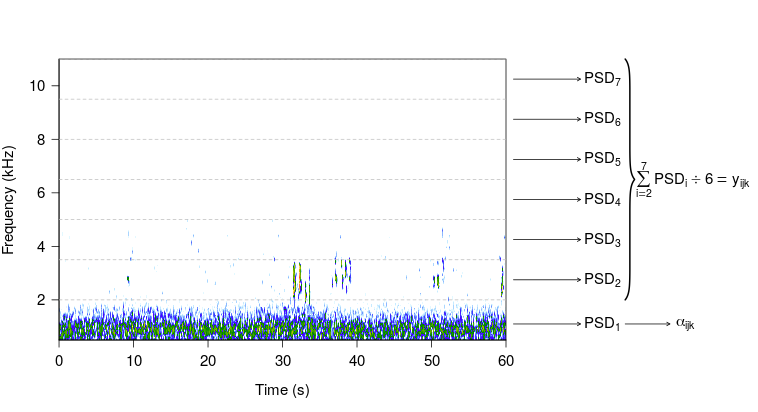}
    \caption{Computation of biological sound ($y_{ijk}$) and anthropogenic noise ($\alpha_{ijk}$) values using the power spectral density for a single recording minute ($i$) at a single location ($j$) at a single time of day ($k$).}
    \label{fig:bioTechno}
\end{figure}

Each one minute soundscape recording was summarized using the PSD as computed by Welch \citep{welch1967}. The PSD represents the amount of soundscape power within each frequency band in units of watts / kHz (Figure \ref{fig:bioTechno}). Each PSD value was vector normalized to account for differences in recorder setting (i.e., system sensitivity, frequency response) and to facilitate comparison of PSD values obtained at different sites \citep{joo2011}. Here we chose to use the Welch spectrum to compute the PSD in units of watts / kHz as opposed to computing the PSD in units of dB re 20 $\mu$Pa$^2$Hz$^{-1}$ \citep{Merchant2015} as this method has been employed in numerous terrestrial soundscape studies to create indices of the amount of biological sounds, anthropogenic noise, and a ratio between the two types of sound \citep{joo2011, gage2017, ndsi, fuller2015, buxton2018, Buxton2018a, Rajan2019}. We computed the PSD for each 1.5 kHz frequency band between 0.5-11 kHz (i.e., a total of seven 1.5 kHz bands), where each value ranged from 0 (no sound) to 1 (filled with sound). 

Anthropogenic noise commonly occurs in the frequency ranges between 0.5 - 2 kHz \citep{ndsi, joo2011, napoletano2004}. Thus, we used the normalized PSD from the 0.5-2 kHz band to represent the amount of anthropogenic noise in each recording following the technique of \cite{ndsi} and the sum of the 1.5 kHz width PSD values from 2 - 11 kHz  to represent the amount of biological sound in each recording. We assume all sound below the frequency of 2 kHz is anthropogenic noise. Such an assumption is generally not safe in western New York because several species communicate in these frequency ranges (e.g., Red Fox (\textit{Vulpes vulpes}), Northern Flicker (\textit{Colaptes auratus}), American Crow (\textit{Corvus brachyrhynchos}), Common Raven (\textit{Corvus corax})). However, the manual analysis found only one common species (American Crow) communicating in this frequency range in our data set, suggesting that using the PSD from the 0.5-2 kHz band will be a fairly reasonable approximation for the total amount of anthropogenic noise in the soundscapes recorded in this study. The anthropogenic and biological sound values were scaled to the range of 0-1 watts/kHz by dividing by the total number of frequency bins involved in their computation (1 and 6, respectively).  We used the \verb+soundecology+ \citep{soundEcology}, \verb+tuneR+ \citep{tuneR}, and \verb+seewave+  \citep{seur2008} packages  within the R Statistical Software \citep{r} environment to compute these measures. PSD values were averaged over the left and right channels to obtain a single value of anthropogenic noise and biological sound for each recording. 

\begin{figure}
    \centering
    \includegraphics[width = 16cm, height = 10cm]{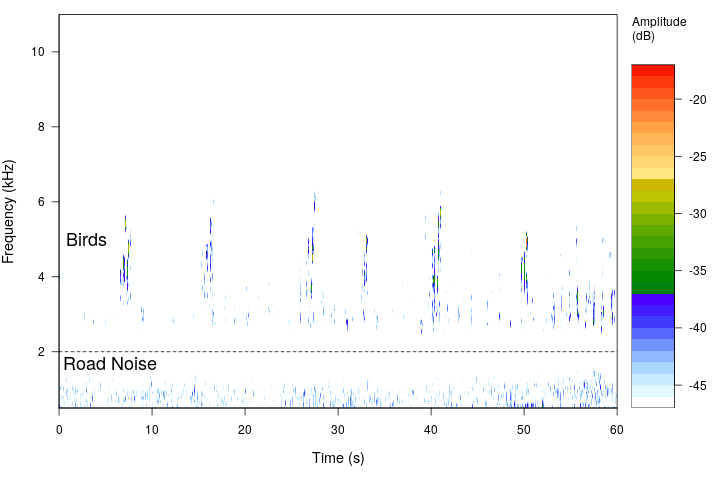}
    \caption{Spectrogram of an example sound recording. The horizontal line represents the cutoff value used to separate anthropogenic and biological sounds.}
    \label{fig:spec}
\end{figure}

\begin{figure}
\centering
\begin{tabular}{cc}
\subcaptionbox{}{\includegraphics[width = 2.5in]{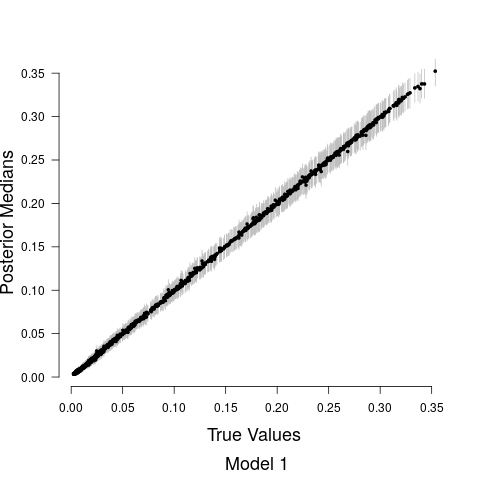}} &
\subcaptionbox{}{\includegraphics[width = 2.5in]{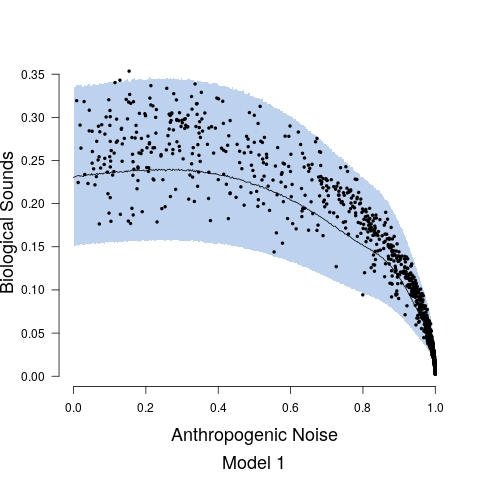}} \\
\subcaptionbox{}{\includegraphics[width = 2.5in]{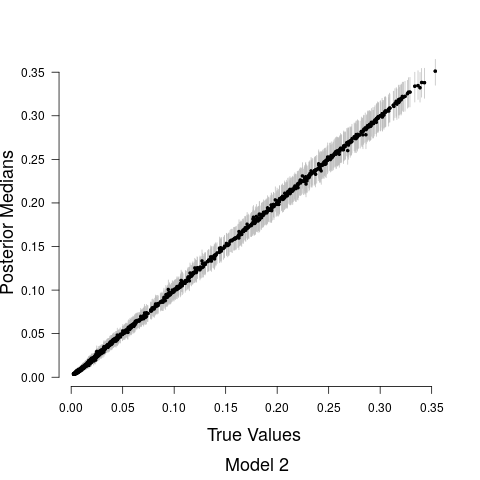}} &
\subcaptionbox{}{\includegraphics[width = 2.5in]{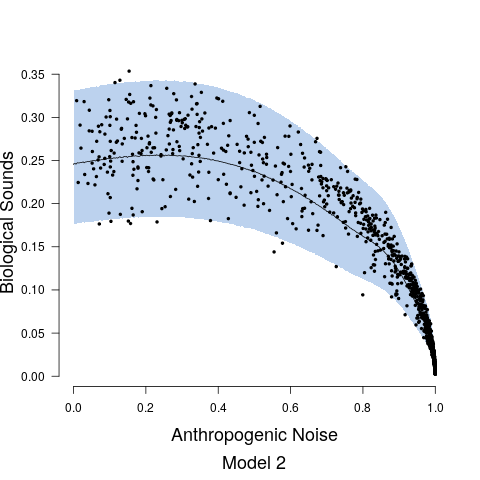}} \\
\subcaptionbox{}{\includegraphics[width = 2.5in]{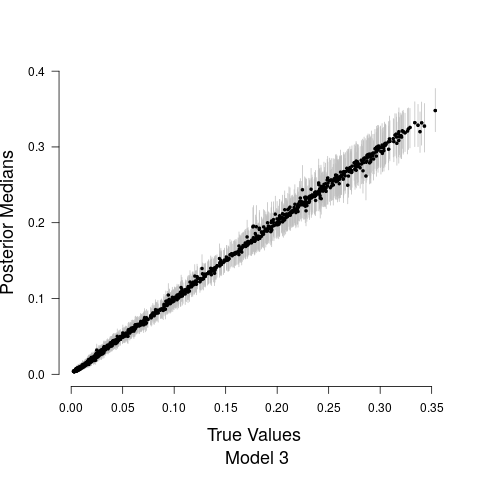}} &
\subcaptionbox{}{\includegraphics[width = 2.5in]{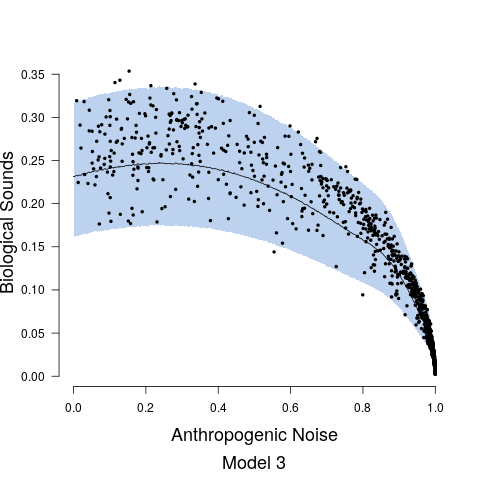}} \\
\end{tabular}
\caption{Stage 2 model fits. 95\% credible intervals are displayed as gray lines in (a), (c), and (e). Posterior medians of model fitted values are displayed in black and 95\% credible intervals are displayed as the blue shaded regions in (b), (d), and (f).}
\label{fig:stage2Fits}
\end{figure}

To further assess our frequency range assumptions for biological and anthropogenic sounds, a subset of recordings was analyzed to determine the proportion of biological sound in the 0.5-2 kHz band. Two one minute recordings for each combination of site and time period were analyzed (108 one minute recordings). The analysis revealed biological sounds were scarcely present in the 0.5-2 kHz band, with a median of 0.9 seconds of each minute containing audible biological sounds (First Quartile: 0 seconds, Third Quartile 3.48 seconds), and a total of 50 recordings containing no audible biological sounds in the 0.5-2 kHz band (Figure~\ref{fig:spec}). The biological sounds within the 0.5 - 2 kHz range also tended to be of low amplitude compared to the anthropogenic noise in the same recording, and thus their impact on the PSD would be minimal. Anthropogenic noise was sometimes present above the 2 kHz region, especially in the 2-3 kHz region. Exploratory data analysis revealed this was in the soundscapes primarily dominated with anthropogenic noise where the anthropogenic noise PSD value was > 0.95. Thus, we acknowledge the values for biological sound at high levels of anthropogenic noise may be overestimated. However, this would not change our results to a high degree, as the biological values at high levels of anthropogenic noise are already quite low (see data points in Figure \ref{fig:stage2Fits}), and so we believe this assumption is valid for these specific soundscapes.

\subsection*{Road Influence}

\begin{figure}
\centering
    \subcaptionbox{}{\includegraphics[width=8cm]{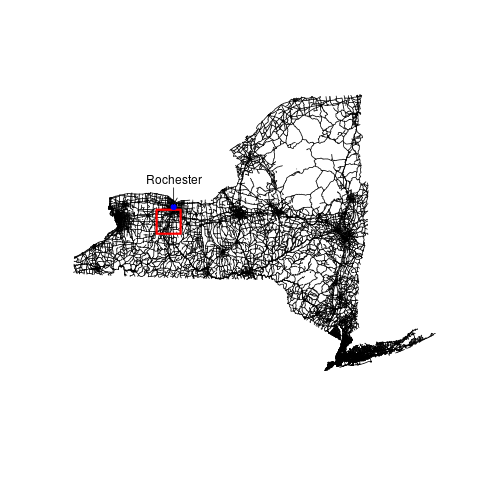}\label{fig:nysRoads}}%
    \subcaptionbox{}{\includegraphics[width=8cm]{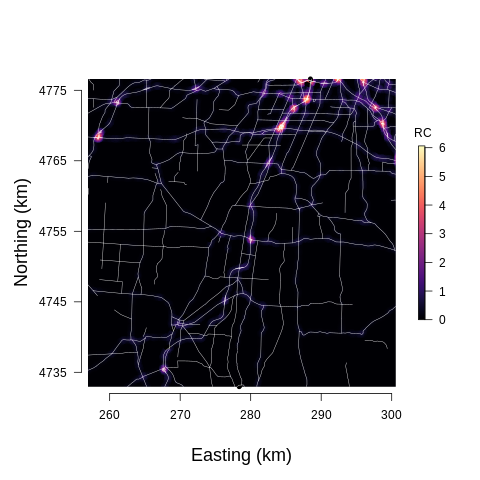}}\label{fig:roadCovariateRoads}%
    \caption{Distribution of roads in New York State. (a) Public road data is displayed across all of NYS. (b) The road covariate is computed at a 30 $\times$ 30 m resolution for the boxed area.}
    \label{fig:roads}
\end{figure}

To assess the influence of roads and traffic sound on the soundscapes we used public data from the New York State GIS Clearinghouse \citep{gisClearinghouse} containing road locations and average speeds. A second data set was acquired from the New York State Department of Transportation \citep{nysDOT} containing the average annual daily traffic (AADT) and an estimate of the percent of trucks on Federal and State highways, and on county and town roads. The roads from these data sets are plotted in Figure \ref{fig:roads}, clearly showing the ubiquitousness of roads throughout  western New York. 
We created a road covariate to quantify the road influence on the soundscape at any given location. This road covariate (RC) took into account 5 factors: 1) average speed; 2) distance of recording site to road; 3) AADT; 4) Percent of trucks; 5) shape of the road. To quantify the shape of the road, we broke each road into 10 $\times$ 10m pixels, obtained the corresponding AADT, speed, and percent of truck values with each road pixel, and computed the distance of each road pixel within 600 m of a given recording site. The 600m boundary is an estimate of how far anthropogenic noise will travel through a forested landscape \citep{forman2000, macLaren2018}. 

We predicted average speed, AADT, and percent of trucks to have a positive relationship with anthropogenic noise and distance to have a negative relationship with anthropogenic noise. Thus, the road covariate is computed as follows for a given 10 $\times$ 10m pixel $i$:

\begin{center}
    $\text{RC}_i = $ AADT + speed + truck - distance
\end{center}

The variables are scaled to provide approximately equal weight to all variables. The complete road covariate for a given recording site is then computed by summing the $\text{RC}_i$ for all locations $i$ within 600m of the given recording site, thus accounting for the shape of the roads near a given recording site. This road covariate is visualized in the study region in Figure \ref{fig:roads}, indicating the covariate is only high near roads, and highest near intersections in the Rochester area. 

Quantification of roads was limited to the roads assessed by the New York State Department of Transportation. These data come primarily from 12,000 annual short traffic counts of 2-7 days of duration. More counts took place in urban areas than in rural and agricultural areas \citep{nysDOT}, which could potentially lead to the road covariate being an underestimate in rural and agricultural regions.

\subsection*{Model}

We seek a model that: 1) provides parameter estimates and associated uncertainty regarding the relationship between biological sound and anthropogenic noise; 2) assesses the amount of anthropogenic noise variance explained by the road covariate; 3) enables biological sound and anthropogenic noise prediction with associated uncertainty. Importantly, we take the view that biological sound is conditional on anthropogenic noise,  and both variables are observed with error. We considered three hierarchical Bayesian models of increasing complexity, henceforth referred to as Model 1, Model 2, and Model 3. Each model consisted of two stages. Stage 1 models anthropogenic noise as a function of the road covariate. Stage 2 models biological sound conditional on Stage 1 such that uncertainty in observed anthropogenic noise is appropriately propagated through the two stages for inferences and subsequent prediction \citep{lunn2013}.

Consider the PSD value for anthropogenic noise $\alpha_{i,j,k}$ and the road covariate $x_{j}$, where $i = 1, 2, \dots , 29$ is the minute of the continuous 29-minute recording, $j = 1, 2, \dots , 18$ indexes recording site, and $k = 1, 2, 3$ indexes time of day. All first stage models use beta regression to account for the bounded support of $\alpha_{i, j, k}$ on $[0, 1]$ and follow the mean and precision parameterization detailed in \cite{ferrari2004}. 

Exploratory data analysis revealed the relationship between anthropogenic noise and the road covariate was non-linear and residuals (i.e., after accounting for the road covariate) were serially correlated with non-constant variance. These features were accommodated using cubic b-splines to obtain a smooth curve over the anthropogenic noise and road covariate functional relationship, and a temporally structured random effect to acknowledge the correlation among the one-minute anthropogenic noise sound bites over each 29 minute recording. More specifically, the random effect followed a multivariate normal distribution with mean 0 and an AR(1) covariance matrix. 

Inferences proceeded by assigning model parameters non-informative prior distributions then a Markov Chain Monte Carlo (MCMC) algorithm sampled from posterior distributions. The full hierarchical model for Model 1, including prior specifications, is detailed below ($[a \mid b]$ is the probability distribution of $a$ conditional on $b$) : \\

\textbf{Stage 1}: 
\begin{align*}
    [\bm{\beta}_{\alpha}, \sigma^2_{\alpha}, \rho_{\alpha}, &\phi_{\alpha}, w_{i, j, k} \mid x_{j}, \alpha_{i, j, k}] \propto \\
    &\prod_{i = 1}^{29} \prod_{j = 1}^{18} \prod_{k = 1}^{3} \space \text{beta}(\alpha_{i, j, k} \mid g(\bm{\beta}_{\alpha}, x_{j}, w_{i, j, k}) \phi_{\alpha}, (1 - g(\bm{\beta}_{\alpha}, x_{j}, w_{i, j, k}))\phi_{\alpha}) \space \times \\
    & \text{multivariate normal}(w_{i, j, k} \mid 0, \sigma^2_{\alpha} \bm{\Sigma}(\rho_{\alpha})) \space \times \\
    & \text{inverse gamma}(\phi_{\alpha} \mid 2, 20000) \space \times \\
    & \text{inverse gamma}(\sigma^2_{\alpha} \mid 2, 5) \space \times \\
    & \prod_{l = 1}^{5} \space \text{normal}(\beta_{\alpha,l} \mid 0, 10000) \space \times \\
    & \text{uniform}(\rho_{\alpha} \mid 0.1, 1)
\end{align*}

where $g(\bm{\beta}_{\alpha}, x_{j}, w_{i, j, k}) = \text{inverse logit}(\bm{Z}_{x_{j}}\bm{\beta}_{\alpha} + w_{i, j, k})$, $\bm{Z}_{x_{j}}$ is the row in the b-spline design matrix for the specific value of the road covariate $x_j$, $w_{i, j, k}$ is the random effect with mean 0 and an AR(1) covariance structure with variance $\sigma^2_{\alpha}$, correlation $\rho_{\alpha}$, and $n \times n$ matrix $\bm{\Sigma}$ (with block diagonal structure where each block is the covariance among 29 consecutive sound bites), $\bm{\beta}_{\alpha}$ are spline regression coefficients, and $\phi_{\alpha}$ is the precision. 

Point and interval estimates for parameters and fitted values were obtained from the joint posterior distribution \citep{gelman04}. Recall, the central role of the Stage 1 model is to explore the relationship between the road covariate and $\alpha$, and propagate the uncertainty in $\alpha$ to the Stage 2 model for the biological sound. This was accomplished by obtaining $M$ post burn-in samples of the fitted values from Stage 1, i.e., $M$ $n \times 1$ vectors of $\hat{\bm{\alpha}}^{(m)}=(\hat{\alpha}^{(m)}_1, \hat{\alpha}^{(m)}_2, \hat{\alpha}^{(m)}_3, \ldots, \hat{\alpha}^{(m)}_n)^\top$ where $m = 1, \dots , M$, and by using them as the covariate in a similar mixed effect beta regression model for the biological sound, $y_{i, j, k}$. The overall structure of Stage 2 is exactly the same as Stage 1, with the exception that each MCMC iteration $m$ fits the biological sound to a different sample $\hat{\bm{\alpha}}^{(m)}$. Stage 2 takes the following form, where all parameters are analogous to Stage 1: \\

\textbf{Stage 2}: 
\begin{align*}
    [\bm{\beta}_{y}, \sigma^2_{y}, \rho_{y}, &\phi_{y}, v_{i, j, k} \mid \hat{\alpha}_{i, j, k}^{(m)}, y_{i, j, k}] \propto \\
    &\prod_{i = 1}^{29} \prod_{j = 1}^{18} \prod_{k = 1}^{3} \space \text{beta}(y_{i, j, k} \mid g(\bm{\beta}_{y}, \hat{\alpha}_{i, j, k}^{(m)}, v_{i, j, k}) \phi_{y}, (1 - g(\bm{\beta}_{y}, \hat{\alpha}_{i, j, k}^{(m)}, v_{i, j, k}))\phi) \space \times \\
    & \text{multivariate normal}(v_{i, j, k} \mid 0, \sigma^2_{y} \Sigma(\rho_{y})) \space \times \\
    & \text{inverse gamma}(\phi_{y} \mid 2, 2000) \space \times \\
    & \text{inverse gamma}(\sigma^2_{y} \mid 2, 2) \space \times \\
    & \prod_{l = 1}^{8} \space \text{normal}(\beta_{y,l} \mid 0, 10000) \space \times \\
    & \text{uniform}(\rho_{y} \mid 0.1, 1).
\end{align*}

While Model 1 does accommodate the serial correlation among the one-minute sound bites, it does not acknowledge within day (i.e., morning, afternoon, and evening) repeated measures aspect of the sampling design. This within day covariance is explicitly taken into account in Model 2 by replacing the scalar variance parameters, $\sigma^2_{\alpha}$ and $\sigma^2_{y}$, with a $3 \times 3$ covariance matrix, $\bm{\lambda}_{\alpha}$ and $\bm{\lambda}_{y}$, whose diagonal elements represent the random effect variance for the respective time period (morning, afternoon, evening) and whose off-diagonal elements represent the covariance between recordings in different time periods. Unlike in Model 1, this structure allows us to make inferences about similarities or differences between the soundscape recordings across the three time periods. The $\bm{\lambda}$'s are modeled with a non-informative inverse-Wishart prior with degrees of freedom $3$ and a diagonal scale matrix with all diagonal elements equal to $0.1$. We use Kronecker products to obtain the desired structure of the covariance matrix, and apply this structure in both Stage 1 and Stage 2.

\begin{figure}
\centering
\begin{tabular}{cc}
\subcaptionbox{}{\includegraphics[width = 2.5in]{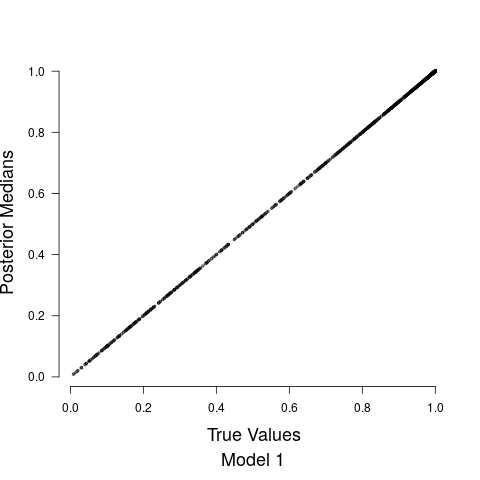}} &
\subcaptionbox{}{\includegraphics[width = 2.5in]{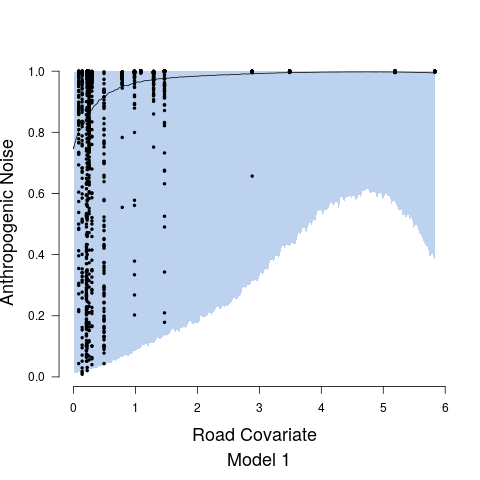}} \\
\subcaptionbox{}{\includegraphics[width = 2.5in]{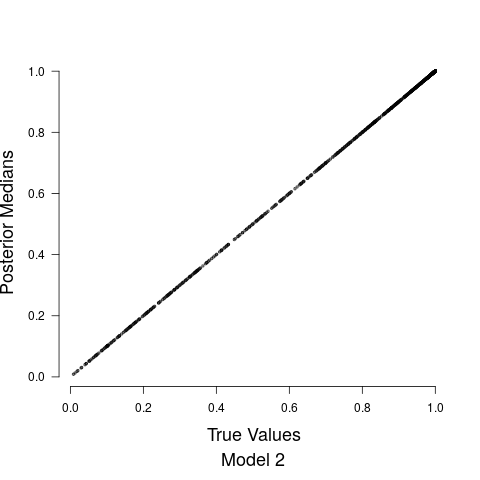}} &
\subcaptionbox{}{\includegraphics[width = 2.5in]{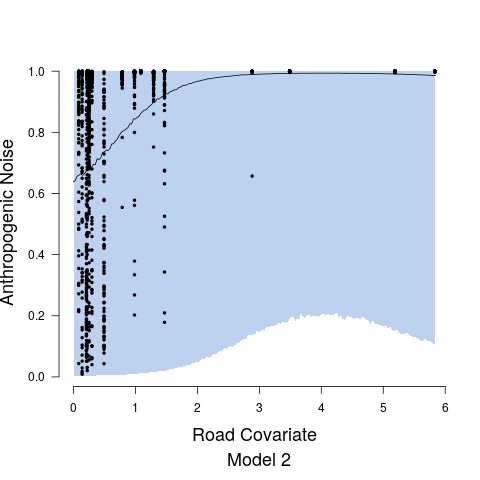}} \\
\subcaptionbox{}{\includegraphics[width = 2.5in]{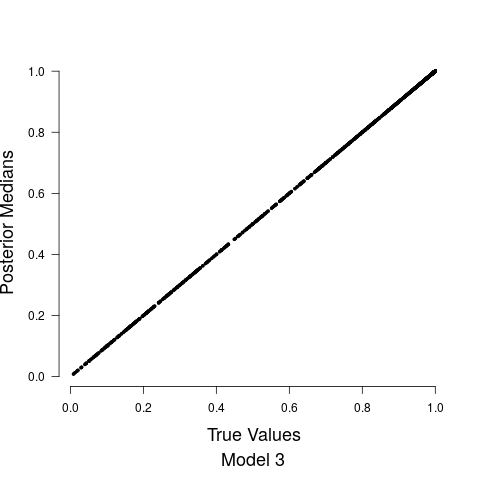}} &
\subcaptionbox{}{\includegraphics[width = 2.5in]{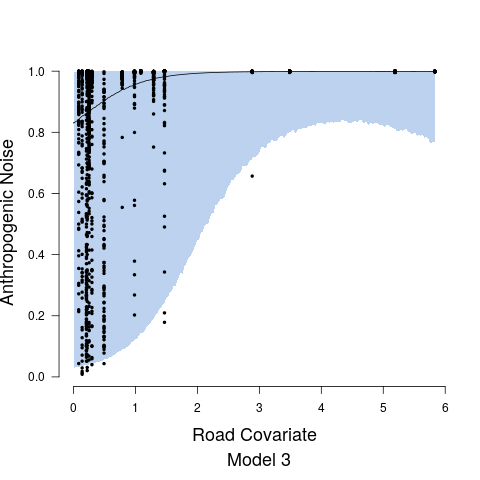}} \\
\end{tabular}
\caption{Stage 1 model fits. 95\% credible intervals are displayed as gray lines in (a), (c), and (e). Posterior medians of model fitted values are displayed in black and 95\% credible intervals are displayed as the blue shaded regions in (b), (d), and (f).}
\label{fig:stage1Fits}
\end{figure}

After examining output from Model 1 and Model 2, diagnostic plots showed observed versus fitted values exhibited heteroskedasticity and associate credible intervals were not appropriately capturing the variability. This non-constant variance was directly addressed in Model 3. For Stage 1, the heteroskedasticity resulted from the relationship between anthropogenic noise and the road covariate. This was remedied by fitting two separate precision parameters $\phi_{\alpha,l}$ and $\phi_{\alpha,u}$ the expression of which was controlled by an indicator function such that $\phi_{\alpha,l}$ is the precision at values of the road covariate less than 2, while $\phi_{\alpha,u}$ is the precision at values of the road covariate greater than 2. While we could have formally estimated the indicator function break point parameter, it was clear from diagnostic plots that a road covariate value of 2 was adequate, see, e.g., Figure~\ref{fig:stage1Fits}.   Both $\phi_{\alpha,l}$ and $\phi_{\alpha,u}$ are modeled with non-informative inverse gamma priors. In Stage 2, we model the precision parameter $\phi_y$ as a function of anthropogenic noise, specifically taking the form $\phi_{y,1} + \phi_2 \text{exp}(\hat{\alpha}_{i, j, k})$.  We modeled $\phi_{y,1}$ and $\phi_{y,2}$ using vague uniform priors from 0 to 10000. 

\subsection*{Prediction}

We seek to develop statistically valid maps of anthropogenic noise and biological sound that reflect the relationships obtained from the three models with associated uncertainty quantification. We computed the road covariate as described previously across a square region in western New York (Figure~\ref{fig:roads}). The posterior predictive distribution for anthropogenic noise is 

\begin{equation}\label{Stage1PP}
    [\bm{\alpha}^* \mid \bm{\alpha}, \bm{x}] = \int_{-\infty}^{\infty} [\bm{\alpha}^* | \bm{\theta}_{\alpha}] [\bm{\theta}_{\alpha} | \bm{\alpha}] d\bm{\theta}_{\alpha}
\end{equation}

where $\bm{\alpha}^*$ is a vector of anthropogenic noise values at new locations, $\bm{x}$ is a vector of road covariate values at new locations, and $\bm{\theta}_{\alpha}$ is a vector of Stage 1 parameters. Similarly, the posterior predictive distribution for biological sound is 

\begin{equation}\label{Stage2PP}
    [\bm{y}^* | \bm{y}, \hat{\bm{A}}] = \int_{-\infty}^{\infty} [\bm{y}^* | \bm{\theta}_{y}] [\bm{\theta}_{y} | \bm{y}] d\bm{\theta}_{y}   
\end{equation}

where $\bm{y}$ is a vector of biological sound values at new locations, $\hat{\bm{A}}$ is an $n^* \times M$ matrix, where $n^*$ is the number of new locations to predict, and $M$ is the number of post-burn MCMC iterations of the fitted values of Stage 1, and $\bm{\theta}_{y}$ is a vector of Stage 2 parameters. 

The integrals in (\ref{Stage1PP}) and (\ref{Stage2PP}) are approximated using MCMC based composition sampling \citep[see, e.g.,][]{banerjee2014}. Posterior predictive samples from $\bm{\alpha}^*$ and $\bm{y}^*$ were used to compute anthropogenic noise and biological sound medians and associated credible intervals. 

\subsection*{Convergence Diagnostics and Model Validation}

Diagnostics were performed to ensure convergence of the MCMC chains. We used a combination of visual assessment of trace plots and an alternative version of the Gelman-Rubin diagnostic that does not assume normality of the correction factor \citep{brooks1998}.

True assessment of the predictive ability of a model requires some form of hold out data that are not used for fitting the model. To accomplish this, we performed a k-fold cross validation technique with $k = 6$ \citep{vehtari2002}. This technique requires fitting the model $k$ times, where each time the model is fit on $n/k$ data points, where $n$ is the length of the data set. Each run of the model fits on a different portion of the data, and predicts the remaining $n - n/k$ hold out values. Since these data are not used in the model fitting process, they represent true draws from the posterior predictive distribution that can be compared with the actual values of the data to assess the predictive capabilities of the model. We used the Continuous Rank Probability Score \citep{gneiting2007} and the Expected Log Pointwise Predictive Density \citep{vehtari2016} to compare the predictive capabilities of the model. Further, we computed the 95\% coverage interval for each of the models, which gives us the percentage of the actual data values that fall within the 95\% credible interval of the model. 

\subsection*{Software Implementation}

MCMC samplers were written in C++ using an Adaptive-Metropolis-within-Gibbs algorithm \citep{roberts2009}. Computationally expensive matrix operations were coded using the Intel Math Kernel Library (MKL) \verb+BLAS+ and \verb+LAPACK+ routines. Prediction and model validation were performed in both C++ and R utilizing the \verb+scoringRules+ package to compute the CRPS \citep{scoringRules}. All subsequent analysis was performed in R \citep{r}. All data and code will be made available on GitHub upon acceptance or upon request.
        
\section*{Results}

Figure~\ref{fig:stage1Fits} shows the relationship between the road covariate and anthropogenic noise depends upon the value of the road covariate. When the road covariate is high, there are large amounts of anthropogenic noise, aligning with intuition and previous research suggesting that anthropogenic noise is higher in more urban areas \citep{pijanowski:2011, pijanowski-2-2011}. However, at low values of the road covariate we obtain essentially no information about anthropogenic noise in the soundscape. Candidate model parameter estimates are given in Table \ref{tab:stage1Params}. Convergence diagnostics suggested rapid convergence for all model parameters with the exception of a few spline coefficients, $\beta$'s, in Stage 1. Such lack of convergence is common in spline-based regression components, especially in the presence of an additive structured random effect \citep{wood2006,hanks2015}. This lack of convergence is of no concern because we are not interested in interpreting the individual spline basis function coefficients---we simply look to Stage 1 to adequately capture the uncertainty in observed anthropogenic noise, and characterize the relationship between anthropogenic noise and the road covariate.

Model fits are shown in Figure~\ref{fig:stage2Fits} along with the estimated relationship between biological sound and anthropogenic noise. Generally, as anthropogenic noise increases, biological sound decreases, aligning with previous research \citep{pijanowski:2011}. Model 3 performed best according to all model validation criteria; however, all models performed very well. Candidate model parameter estimates are given in Table \ref{tab:stage2Params}. All Stage 2 model parameters showed strong convergence.

To ease interpretation, covariance matrix estimates are often best expressed as correlations. Converting each MCMC sample from the $\bm{\lambda}$'s posterior to a correlation provides access to the corresponding correlation matrix posterior, which is summarized in Tables~\ref{corStage1} and \ref{corStage2} for Stage 1 and 2, respectively.   

Because inference is primarily focused on estimating biological sound given anthropogenic noise in the soundscapes, we perform model comparison only for Stage 2 models. A 6-fold-cross validation was used to compare candidate models' out-of-sample prediction using the CRPS and ELPD. High values of the ELPD and low values of the CRPS suggest a better model fit. We also report the percentage of points covered by the 95\% credible intervals of the predicted biological sound versus anthropogenic noise relationship, which should ideally cover 95\% of the data points (Table~\ref{tab:modelComparison}).

\begin{figure}
\centering
\begin{tabular}{cc}
\subcaptionbox{}{\includegraphics[width = 2.5in]{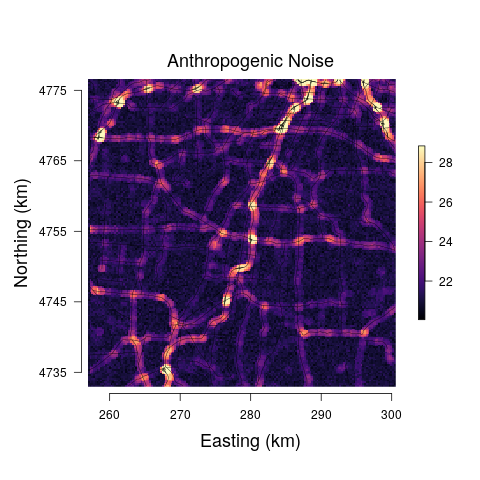}} &
\subcaptionbox{}{\includegraphics[width = 2.5in]{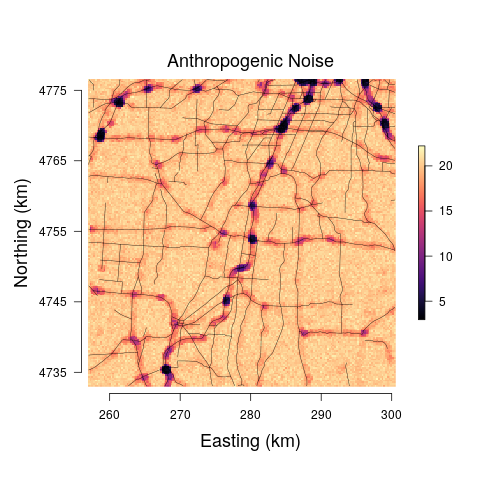}} \\
\subcaptionbox{}{\includegraphics[width = 2.5in]{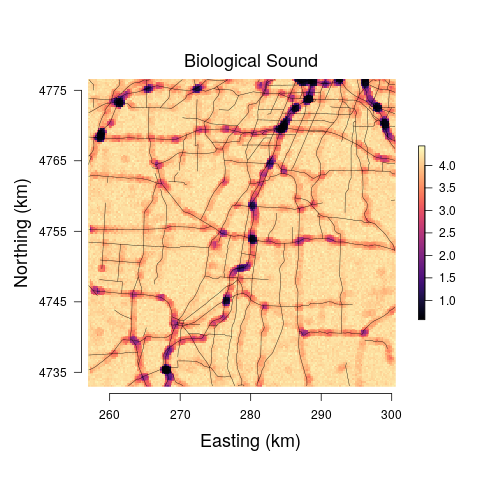}} &
\subcaptionbox{}{\includegraphics[width = 2.5in]{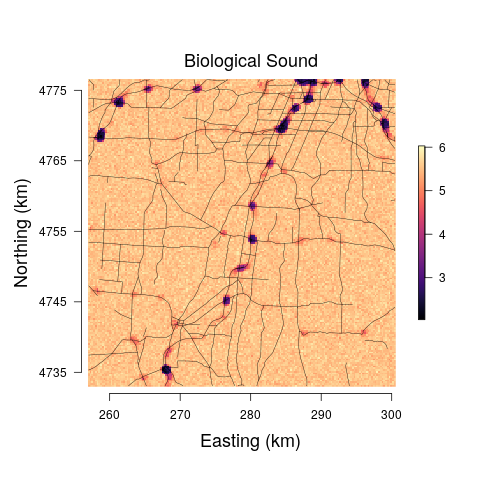}} \\
\end{tabular}
\caption{Model 3 afternoon predictions of anthropogenic noise and biological sound over a sample region in western New York. Posterior medians are shown in (a) and (c), while posterior 95\% credible interval widths are shown in (b) and (d).}
\label{fig:predictions}
\end{figure}

The models yield anthropogenic noise and biological sound prediction at the 29 minute observation resolution for three times of the day. Such fine temporal resolution is likely not that useful from an assessment or management perspective. Hence, we summed each 29 minute biological sound and anthropogenic noise posterior predictive sample, resulting in a posterior predictive distribution for the total anthropogenic noise and biological sound at each pixel across the study area for morning, afternoon, and evening. The median and range between the upper and lower 95\% credible interval bounds for each pixel-level predictive distribution were mapped. Very little differences were detected among the models and between predictions at the morning, afternoon, and evening, and thus we only present posterior predictive maps for the afternoon soundscapes for Model 3 in Figure~\ref{fig:predictions}. 

\section*{Discussion}

There is recent widespread recognition that anthropogenic traffic noise alters the acoustic habitat for many animal species \citep{barber:2011, barber:2010, buxton:2017}, and influences biodiversity within reach of that sound \citep{katti2004, mcclure2013}. Acoustic soundscape recordings are now commonly used in the field of ecoacoustics to monitor natural soundscapes impacted by anthropogenic noise \citep{pijanowski:2011, farina2018}. An important and complex challenge in ecoacoustics is to determine the impacts of road noise on biodiversity. As a potential first step to address this problem, we proposed a novel hierarchical Bayesian model to explore the relationships between public road data, anthropogenic noise, and biological sounds. Specifically, we developed three two-stage mixed effects beta regression models to assess the degree to which public traffic data explains variability in anthropogenic noise and to characterize the relationship between biological sound and anthropogenic noise in western New York soundscapes. The models were compared using inference delivered and out-of-sampled prediction. Models were then applied to provide anthropogenic noise and biological sound predictive maps over a sample region in western New York using public road data. The predictive maps have large uncertainty in locations far from roads, resulting from the temporal coarseness of public road data used as a proxy for traffic noise. This suggests a need for finer temporal resolution traffic noise data. The use of such data, in combination with the proposed modeling framework, could have important implications for the development of soundscape and sound management policies.

Road noise is one of the most ubiquitous sources of anthropogenic noise \citep{projectRemote, barber:2010}. In Stage 1 of our proposed modeling framework, we model the relationship between public road traffic data and anthropogenic noise to determine how well the public traffic data explains variability in anthropogenic noise (Figure \ref{fig:stage1Fits}). Model results ultimately displayed the inability of the road covariate to explain variability in anthropogenic noise at low values. The large uncertainty in the relationship between anthropogenic noise and the road covariate is also evident in the predictions of anthropogenic noise given new values of the road covariate, as the credible interval widths are extremely large at areas where the road covariate is low (Figure~\ref{fig:predictions}). The large variation in the anthropogenic noise values at low levels of the road covariate could potentially be improved by incorporating important variables regarding the road surface type that likely influence the propagation of road sound throughout the environment, which were not available in the data sets used in this study. However, the data showed high variability in the anthropogenic noise values within each site, suggesting it is more likely that high variability in this relationship is a result of individual effects that are not accounted for by site level variables. These individual effects are likely a result of large variations in the number/type of automobiles on the road at any given minute of time. We listened to all recordings, and confirmed road sound was the most prominent source of anthropogenic noise, further suggesting the high variation of the relationship between the road covariate and the human sound is a result of high temporal variation in the number of cars on a given road, a phenomenon that is well-described in literature on traffic sound modeling \citep{can2008, bert2005}. The use of models that incorporate the dynamic temporal changes of road sound across time could help account for the temporal changes in traffic and subsequent traffic sound if traffic data are limited as in this study \citep{can2008}.  Utilizing crowd-sourced traffic data from traffic and navigation apps (i.e., Google Maps, Waze) is an intriguing alternative that would enable more time-specific measures of traffic and subsequently the sound it produces. Such space-time data, in combination with the modeling frameworks proposed here, could result in near real-time maps of anthropogenic noise that could have important implications for the development of sound management policies.

Using spatially or temporally structured random effects in soundscape models can improve model accuracy when the data are limited and the researcher suspects there are individual effects causing variation not explained by the data \citep{clarkTextbook}. In this study, the use of random effects allowed us to incorporate  temporal dependence between recordings, obtain accurate model fits, and predict anthropogenic noise and biological sound despite using a predictor (the road covariate) that does not explain large amounts of variation of anthropogenic noise. 

Understanding the relationship between biological sounds and anthropogenic noise is an important initial step in determining the impact of anthropogenic noise on biodiversity. In Stage 2 of our modeling framework, we quantify the complex, non-linear relationship between biological sound and anthropogenic noise (Figure \ref{fig:stage2Fits}). We modeled this relationship using three candidate models, each increasing in complexity. Table \ref{tab:modelComparison} shows that Model 3 has the highest ELPD, the lowest CRPS values, and the most accurate 95\% coverage. This suggests that accounting for the repeated measures across time of day in the soundscape recordings as well as the non-constant variance in the data provides an improvement in the model.

Soundscapes are highly correlated across both space and time \citep{pijanowski:2011, mullet2016} and thus a modeling framework that quantifies such correlations are important inferential tools. The additional time of day correlation estimates in Models 2 and 3 provide inference on the relationship between the soundscapes over the morning, afternoon, and evening recordings. For Model 3, Stage 1 (Table \ref{tab:cor}), we see the correlation between the random effects of the afternoon recordings with both the random effects of the morning and evening recordings are not different from 0 (i.e., 0 is contained within the 95\% credible interval), whereas the correlation between morning and evening random effects are small but different from 0, with a posterior median of 0.18. This suggests that variations in anthropogenic noise not explained by the road covariate are similar in the morning and evening recordings, although the correlation coefficient of 0.18 suggests this is not a strong relationship. For Model 3, Stage 2 (Table \ref{tab:cor2}), we see similar results in that the correlation between morning and evening recordings is different from 0, with a posterior median of 0.27, suggesting that variations in biological sound not explained by anthropogenic noise are more similar in the morning and the evening recordings than they are between the afternoon recordings and either the morning or evening recordings. This is likely a result of the dawn and dusk choruses, which are captured by the morning and evening recording time periods, respectively. Thus, we see that Model 3 provides slight improvements in terms of the model validation criteria, in addition to providing insights into the temporal relationships between biological sound and anthropogenic noise that are not available from the more simple Model 1. 

To illustrate how our modeling framework can be used to predict anthropogenic and biological sounds over larger spatial areas, we provide soundscape maps of a sample region in western New York at a 250 $\times$ 250 m resolution where we predict anthropogenic noise and biological sound from public road data. Visualization of the posterior median suggests that biological sound is highest in areas farther away from roads, while anthropogenic noise is high in regions of more concentrated and highly used roads. This aligns with previous research and intuition, as the probability of detection of avian species vocalizations is lower closer to roads \citep{parris2008} and  anthropogenic noise increases with the degree of urbanization \citep{pijanowski-2-2011}. However, a visualization of the 95\% credible interval widths shows that there are large amounts of uncertainty associated with these estimates at areas with low anthropogenic noise, largely a result of the inability of anthropogenic noise to be accurately predicted at low levels of the road covariate. Thus, inference drawn from these maps is limited due to our lack of certainty.

Despite the fact that there is a clear negative relationship between biological sound and anthropogenic noise, we see that past a given distance from the road the predictions of biological sound are all very similar. In this study, we were solely interested in the relationship between anthropogenic noise and biological sound and the ability of public road data alone to serve as a predictor. If more accurate predictions of biological sound are desired, it will be important to include covariates in the model that quantify the landscape structures that determine the types of organisms communicating in the soundscape \citep{pijanowski:2011, gage2017}. One example of  successful soundscape maps of biological sound and anthropogenic noise was shown in a study of south-central Alaska from numerous landscape and anthropogenic measures, such as distance to rivers, distance to wetlands, aspect, and snowmobile activity \citep{mullet2016}. In the landscape we have mapped, the habitat ranges from small patches of forest, to agricultural fields, small towns and villages, and suburban development. This range of habitats would be expected to support different assemblages of acoustically communicating species resulting in different biological sound. Accounting for these differences in landscape through the appropriate covariates could also better inform the radius used to calculate the road covariate, as the attenuation and transmission of sound is highly dependent on the structure of the landscape \citep{Royle2018}. In this study we did not have data on important landscape variables and so a 600m boundary was used as an estimate for how far anthropogenic noise will travel through a forested landscape based on \cite{forman} and \cite{macLaren2018}.

The PSD and acoustic indices derived from it (NDSI) have previously been shown to correlate positively with anthropogenic activity \citep{fairbrass2017} and change with landscape structure \citep{fuller2015}. Our soundscape maps support these findings as the PSD of the 0.5-2 kHz range that represents anthropogenic noise is highest in areas of high road concentration. However, the use of the PSD to represent anthropogenic noise and biological sound as we did in this study is limited in application to long-term soundscape monitoring studies. Depending on the location and time of day, numerous organisms communicate within the 0.5-2 kHz region that is designated as anthropogenic noise. In our study, a manual analysis of the recordings revealed only one species commonly singing within the 0.5-2 kHz region, supporting the use of the PSD values as proxies of biological sound and anthropogenic noise in this setting. In addition, we only recorded on days with no rain and minimal wind, minimizing the presence of abiotic sounds in the soundscape recordings, which is often not possible over vast spatio-temporal regions. Analysis of long-term monitoring of soundscapes where such assumptions are not valid requires alternative methods to distinguish between biological sound, anthropogenic noise, and abiotic sound. Convolutional neural networks have recently been utilized in a deep learning system called CityNet to predict the presence or absence of biological sound and anthropogenic noise in urban soundscapes \citep{fairbrass2019}. Recent work using the spectral properties of sound as is done in Music Information Retrieval also shows promise for distinguishing between the three soundscape components \citep{bellisario2019, bellisario2019-2}. Further developing these methods, in conjunction with current acoustic indices and landscape measurements, could provide reasonable estimations of the relative amounts of biological sound, anthropogenic noise, and abiotic sound in a soundscape to allow for long-term monitoring of soundscapes and landscape health. 

Like many ecological studies, data collection was limited as a result of resource availability (i.e., number of acoustic recorders). In this study, each soundscape was recorded on a single day between May and June for a total of 90 minutes. While this data paucity could have potential implications on inference, exploratory data analysis revealed that the specific day of the recording and time from sunrise did not explain large amounts of variability in the relationships between the road covariate, anthropogenic noise, and biological sound, suggesting we are accurately capturing the relationship between anthropogenic and biological components in these soundscapes. Thus, we believe the available data and analysis provide valid inference about the soundscape dynamics in our study system and allow us to adequately explore the study objectives. Further, our proposed modeling framework is broadly applicable to settings where monitoring networks have additional data. 

The proposed models were used to assess the extent to which available traffic data explains variability in anthropogenic noise and to quantify the functional relationship between anthropogenic noise and biological sound. Roads represent the dominant source of anthropogenic noise across the landscape in our study area, and have a large and growing impact around the world \citep{buxton:2017, barber:2011}. Understanding and predicting the sound impacts of roads on biological communities is an important focus of ecoacoustics researchers in many locations \citep{forman2000, herraraMontes2011, mullet2016}. The hierarchical Bayesian framework allows us to obtain parameter estimates, fitted values, and predicted values at new locations all within the same modeling framework. This framework can incorporate a range of soundscape data to explore a variety of topics, such as the relationship between biological sounds and anthropogenic impacts like road sound or habitat fragmentation, the monitoring of species density and population estimates using acoustic recordings, and the recovery of environments to natural/anthropogenic disturbances. Ecologists, conservation biologists, urban planners, and road engineers all have an interest in these questions. Utilization of such a broadly applicable modeling framework will greatly improve our ability to make inference regarding the ways anthropogenic noise contributes to the soundscape and influences biological sound and the biodiversity that it represents. 

\bibliographystyle{apa}
\bibliography{references}

\section*{Tables and Figures}

\begin{table}[h!] 
  \begin{center}
  \caption{Stage 1 posterior parameter medians and 95\% credible intervals, 50\% (2.5\%, 97.5\%). Subscript in parentheses on $\lambda_\alpha$ indicate the row and column element in $\bm{\lambda}_\alpha$.}
  \label{tab:stage1Params}
  \begin{tabular}{c c c c}
    \toprule
    Parameter & Model 1 & Model 2 & Model 3 \\
    \midrule
    $\beta_{\alpha,0}$ & 1.12 (1.10, 1.15) & 0.64 (0.60, 0.65) & 1.59 (1.57, 1.61)\\
    $\beta_{\alpha,1}$ &  3.71 (3.68, 3.74) & 1.01 (0.93, 1.06) & 2.46 (2.36, 2.54)\\
    $\beta_{\alpha,2}$ & 3.84 (3.59, 4.05) & 5.38 (5.00, 5.86) & 7.40 (7.00, 7.59) \\
    $\beta_{\alpha,3}$ & 6.83 (6.60, 7.41) & 4.81 (4.45, 5.18) & 6.81 (6.06, 7.45) \\
    $\beta_{\alpha,4}$ & 4.37 (4.19, 5.28) & 4.12 (3.90, 4.46) & 6.44 (6.23, 6.63) \\
    $\phi_{\alpha}$ & 232161 (80716, 878895) &  348648 (105344, 738622) & - \\
    $\phi_{\alpha,l}$ & - & - & 92623 (37579, 228416) \\
    $\phi_{\alpha,u}$ & - & - & 239221 (68653, 1976045) \\
    $\lambda_{\alpha,(1,1)}$ & - & 2.85 (2.58, 3.21) & 2.23 (2.10, 2.54) \\
    $\lambda_{\alpha,(2,1)}$ & - & 0.07 (-0.22, 0.36) & 0.02 (-0.21, 0.25) \\
    $\lambda_{\alpha,(3,1)}$ & - &  0.62 (0.29, 0.95) & 0.46 (0.21, 0.73) \\
    $\lambda_{\alpha,(2,2)}$ & - & 3.28 (2.96, 3.73) & 2.63 (2.39, 2.92) \\
    $\lambda_{\alpha,(3,2)}$ & - & 0.11 (-0.22, 0.44) & 0.06 (-0.19, 0.32) \\
    $\lambda_{\alpha,(3,3)}$ & - & 3.78 (3.39, 4.29) & 3.01 (2.73, 3.36) \\
    $\sigma^2_{\alpha}$ & 8.38 (7.04, 10.23) & - & - \\
    $\rho_{\alpha}$ & 0.89 (0.87, 0.91) & 0.92 (0.90, 0.94) & 0.87 (0.85, 0.90) \\
    \bottomrule
  \end{tabular}
  \end{center}
\end{table}

\begin{table}[h!] 
  \begin{center}
  \caption{Stage 2 posterior parameter medians and 95\% credible intervals, 50\% (2.5\%, 97.5\%). Subscript in parentheses on $\lambda_y$ indicate the row and column element in $\bm{\lambda}_y$. }
  \label{tab:stage2Params}
  \begin{tabular}{c c c c}
    \toprule
    Parameter & Model 1 & Model 2 & Model 3 \\
    \midrule
    $\beta_{y,0}$ & -1.16 (-1.26, -1.09) & -1.14 (-1.20, -1.11)& -1.21 (-1.26, -1.06) \\
    $\beta_{y,1}$ &  -1.08 (-1.16, -0.96) & -1.01 (-1.06, -0.96)& -1.07 (-1.18, -0.98) \\
    $\beta_{y,2}$ & -1.01 (-1.08, -0.95) & -0.99 (-1.06, -0.91) & -1.07 (-1.12, -0.98) \\
    $\beta_{y,3}$ & -1.75 (-1.78, -1.71) & -1.71 (-1.76, -1.65) & -1.74 (-1.78, -1.70) \\
    $\beta_{y,4}$ & -1.87 (-1.91, -1.80) & -1.84 (-1.88, -1.78) & -1.91 (-1.95, -1.85) \\
    $\beta_{y,5}$ & -2.99 (-3.03, -2.96) & -2.98 (-3.02, -2.91) & -3.00 (-3.04, -2.96) \\
    $\beta_{y,6}$ & -3.49 (-3.54, -3.43) & -3.48 (-3.52, -3.43) & -3.54 (-3.59, -3.48) \\
    $\beta_{y,7}$ & -5.04 (-5.09, -4.99) & -5.06 (-5.11, -4.98) & -5.03 (-5.10, -5.00) \\
    $\phi_{y}$ & 7487 (5658, 9606) & 6919 (5526, 9085) & - \\
    $\phi_{y,1}$ & - & - &  128.68 (4.53, 542.99) \\
    $\phi_{y,2}$ & - & - & 1343.38 (1072.11, 1596.93) \\
    $\lambda_{y,(1,1)}$ & - & 0.21 (0.19, 0.23) & 0.21 (0.19, 0.24) \\
    $\lambda_{y,(2, 1)}$ & - & -0.004 (-0.03, 0.02)& -0.01 (-0.04, 0.02) \\
    $\lambda_{y,(3, 1)}$ & - & 0.05 (0.03, 0.08) & 0.06 (0.03, 0.09) \\
    $\lambda_{y,(2, 2)}$ & - & 0.20 (0.18, 0.22) & 0.20 (0.18, 0.23) \\
    $\lambda_{y,(3, 2)}$ & - & 0.007 (-0.02, 0.03) & 0.02 (-0.02, 0.05) \\
    $\lambda_{y,(3, 3)}$ & - & 0.21 (0.19, 0.24) & 0.21 (0.19, 0.24)  \\
    $\sigma^2_y$ & 0.06 (0.05, 0.07) & - & - \\
    $\rho_y$ & 0.82 (0.78, 0.86) & 0.74 (0.70, 0.79) & 0.82 (0.76, 0.86) \\
    \bottomrule
  \end{tabular}
  \end{center}
\end{table}

\begin{table}[h!] 
    \begin{center}
        \caption{Model 3 Stage 1 random effect correlation matrix posterior medians and 95\% credible intervals, 50\% (2.5\%, 97.5\%). Boldface indicates parameter values not containing 0 in the associated 95\% credible interval.}\label{corStage1}
        \label{tab:cor}
        \begin{tabular}{c c c c}
        \toprule
        & Morning & Afternoon & Evening \\
        \midrule
        Morning & - & - & - \\
        Afternoon & 0.01 (-0.08, 0.10) & - & - \\
        Evening & $\bm{0.18 (0.08, 0.28)}$ & 0.02 (-0.07, 0.11) & - \\
        \bottomrule
        \end{tabular}
    \end{center}
\end{table}

\begin{table}[h!] 
    \begin{center}
        \caption{Model 3 Stage 2 random effect correlation matrix posterior medians and 95\% credible intervals, 50\% (2.5\%, 97.5\%). Boldface indicates parameter values not containing 0 in the associated 95\% credible interval.}\label{corStage2}
        \label{tab:cor2}
        \begin{tabular}{c c c c}
        \toprule
        & Morning & Afternoon & Evening \\
        \midrule
        Morning & - & - & - \\
        Afternoon & -0.06 (-0.20, 0.09) & - & - \\
        Evening & $\bm{0.27 (0.14, 0.45)}$ & 0.08 (-0.12, 0.22) & - \\
        \bottomrule
        \end{tabular}
    \end{center}
\end{table}

\begin{table}[h!] 
  \begin{center}
  \caption{Comparison of ELPD, CRPS, and 95\% Coverage Intervals}
  \label{tab:modelComparison}
  \begin{tabular}{c c c c}
    \toprule
     & Model 1 & Model 2 & Model 3 \\
    \midrule
    ELPD & 730.5 & 728.17 & 733.17\\
    CRPS & 0.0098 & 0.0097 & 0.0097 \\
    95\% Coverage & 95.66 & 92.85 & 94.96 \\
    \bottomrule
  \end{tabular}
  \end{center}
\end{table}

\end{document}